\documentclass[10pt]{iopart}
\usepackage{amssymb, amscd,amsfonts,iopams,graphicx,hyperref}
\usepackage[utf8]{inputenc}
\usepackage[T1]{fontenc}
\hypersetup{citecolor=magenta,colorlinks=true,urlcolor=blue}

\expandafter\let\csname equation*\endcsname\relax
\expandafter\let\csname endequation*\endcsname\relax
\usepackage{amsmath}

\begin{document}

\newcommand{\bra}[1]    {\langle #1|}
\newcommand{\ket}[1]    {\left|#1 \right\rangle}
\newcommand{\braket}[2]{\left\langle#1|#2\right\rangle}
\newcommand{\ketbra}[2]{|#1\rangle\!\langle#2|}
\renewcommand{\tr}[1]    {{\rm Tr}\left[ #1 \right]}
\newcommand{\av}[1]    {\langle #1 \rangle}
\renewcommand{\mod}[1]    {\left| #1 \right|}
\newcommand{\modsq}[1]    {\left| #1 \right|^2}
\newcommand{\modsqs}[1]    {\left| #1 \right|^2}

\title[Bell correlations in a vicinity of a quantum critical point]{Bell correlations of a thermal fully-connected spin chain in a vicinity of a quantum critical point}
\author{D A Hamza and J Chwede\'nczuk }
\address{Faculty of Physics, University of Warsaw, ul. Pasteura 5, PL-02-093 Warsaw, Poland}
\ead{jan.chwedenczuk@fuw.edu.pl}
\begin{abstract}
  
  Bell correlations are among the most exotic phenomena through which quantum mechanics manifests itself. Their presence signals that the system can violate the postulates of local realism, once believed to
  be the nonnegotiable property of the physical world. The importance of Bell correlations from this fundamental point of view is even straightened 
  by their applications---ranging from quantum cryptography through quantum metrology to
  quantum computing. Hence it is of growing interest to characterize the ``Bell content'' of complex, scalable many-body systems. 

  Here we perform the detailed analysis of the character and strength of many-body Bell correlations in interacting multi-qubit systems with particle-exchange symmetry. Such configuration can be mapped onto an effective Schr\"odinger-like equation, which allows for precise analytical predictions. We show that in the vicinity of the quantum critical point, these correlations
  quickly become so strong that only a fraction of qubits remains uncorrelated. We also identify the threshold temperature, which, once overpassed, empowers thermal fluctuations that destroy
  Bell correlations in the system. 

  We hope that the approach presented here, due to its universality, could be useful for the upcoming research on genuinely nonclassical Bell-correlated complex systems. 
\end{abstract}
\maketitle

\section{Introduction}

The Bell's rebuke of the Einstein-Podolsky-Rosen (EPR) call for complementing quantum mechanics with a local and realistic theory~\cite{epr} was conceived with the least complex but still 
highly non-trivial many-body configuration: two spin-$1/2$ particles~\cite{Bell}. Though its deceptive simplicity, for many years this system has been driving the progress
of  both the theory and the experiment. In 1969, the Clauser-Horne-Shimony-Holt inequality~\cite{chsh} that refined the original Bell's formulation opened the way for
first experiments showing the violation of Bell inequalities~\cite{test1,test2,test3,test4,test5,test6,test7,test8,test9,test10,test11,test12}. 
Finally, loophole-free tests of Bell nonlocality were reported, discarding local realism in quantum mechanics for good (apart from other possibilities like the global realism~\cite{loophole}, or some hard to exclude, though highly unlikely and unreasonable
scenarios, like a global conspiracy of producers of microchips or detection equipment, intended to fool the experimentalists).
The significance of these efforts was acknowledged by the Nobel committee and the prize was awarded in 2022.

With the much descried arrival of the noisy intermediate-scale quantum (NISQ) devices, the research interest has shifted towards more complex systems. It has become of growing relevance to understand how
the fundamental quantum relations, such as the entanglement~\cite{sch,horodecki2009entanglement}, the EPR steering~\cite{epr,steering3,yadin2021metrological} 
and the Bell nonlocality can be created and detected, in particular in multi-qubit systems. 

The problem of detecting many-body Bell correlations, which are the main focus of this work, has been formulated for $N$-qubit systems using $N$-body correlation functions 
in various ways, most notably by means of Mermin-Bell inequalities~\cite{mermin1990extreme}, and finally expressed in the all-encompassing form~\cite{zukowski2002bell}. 
This general formulation has different variants (i.e., forms of correlation functions, adapted for a concrete physical case) and one particularly useful is the one 
formulated in Refs.~\cite{he2010bell,he2011entanglement,steering2}. It enabled to establish the link between the Bell nonlocality and quantum-enhanced metrology~\cite{PhysRevLett.126.210506}
or determine the strength of many-body Bell correlations in a commonly used method of creating atomic spin-squeezed states~\cite{kitagawa1993squeezed,wineland1994squeezed}, 
i.e., the one-axis twisting method~\cite{esteve2008squeezing,appel2009mesoscopic,pezze2009entanglement,Treutlein2010,Oberthaler2010,berrada2013integrated,interferReview}.

Here we scrutinize the emergence of Bell correlations in the versatile and scalable multi-qubit system. It is a collection of $N$ qubits 
where each pair interacts with the same strength. Such setup can be realized using different platforms well-suited for quantum technologies, such as 
the fully-connected 
Ising spin-chain~\cite{Blatt:2012aa,zhang2017observation,RevModPhys.93.025001,joshi2022observing,morong2021observation,dumitrescu2022dynamical,feng2023continuous} 
or alternatively by trapping the Bose-Einstein condensate in a double-well potential.
While throughout this work we will refer to the latter case, it is equivalent, from the perspective of problems considered here, to the former.
In this scenario, the separated single-body states localized around the two minima of the trap play the role of the pair of single-qubit levels. In the tight-binding regime, this system
is described by the Bose-Hubbard (BH) Hamiltonian, where the on-site two-body interactions compete with the 
coherent Josephson tunneling across the barrier of the double-well potential~\cite{jo2007long,Hofstetter_2018,PhysRevLett.129.090403,schumm2005matter,PhysRevA.107.013311,2017,gati2006noise}. 
On the attractive side of the interaction, there exists a quantum phase transition 
(QPT)~\cite{dziarmaga2002dynamics,trenkwalder2016quantum,PhysRevA.78.042106,PhysRevA.78.042105,PhysRevA.90.022111,PhysRevE.93.052118,PhysRevLett.124.120504,PhysRevLett.123.170604,Guhne_2005,Toth_2014,
PhysRevResearch.5.013158}, 
upon passing of which the properties of the BH Hamiltonian suddenly change---from a gaussian state to the macroscopic
superposition~\cite{PhysRevA.78.042106,PhysRevA.78.042105,PhysRevA.90.022111,PhysRevE.93.052118,PhysRevLett.124.120504,PhysRevLett.123.170604}. We demonstrate that in the vicinity of this point, the Bell correlations are genuinely many-body, i.e., a macroscopic fraction of qubits is Bell-correlated. 
To show this, we derive a simple yet informative formula, that allows to predict the strength of Bell correlations related to this emergent macroscopic superposition. 
Next, we consider non-zero temperatures and analytically demonstrate that the relevant temperature scale, above which the thermal fluctuations ``kill'' the Bell nonlocality, is related
to the energy gap between the ground- and the first-excited state. This allows to derive a clear criterion for the critical---from the point of view of many-body Bell correlations---temperature.

We hope that the presented analysis will contribute to the emerging field of the NISQ devices, allowing for precise characterization of multi-qubit states and for the accurate planning of future experiments.

\section{Many-Body Quantum Systems}
Consider a collection of $N$ qubits (spin-$1/2$ particles) fully connected via the distance-independent two-body interaction of strength $U$. The spins are subject to an external uniform magnetic field aligned
by the $x$-axis, which amplitude is proportional to $\Omega$. One possibility of modelling such system is via the Ising model, the one on which we focus in this work. The Hamiltonian reads
\begin{align}\label{eq.ham.ch}
  \hat{H}=-\Omega\sum_{i=1}^{N}\hat{\sigma}_{x}^{(i)}+\frac U2\sum^{N}_{i\neq j=1} \hat{\sigma}_{z}^{(i)}\hat{\sigma}_{z}^{(j)}.
\end{align}
Here  $\hat{\sigma}_{\xi}^{(k)}$ is the $\xi$-component of the triad of Pauli matrices for the $k$-th spin (i.e., $\xi=x,y,z$).
Note that this Hamiltonian is invariant under the exchange of any pair of spins, hence all its eigen-states posses this symmetry. 
Thus to analyze the many-body properties of the ground- and the thermal-states, it is convenient to map it onto the bosonic BH Hamiltonian in the tight-binding limit, namely
\begin{align}\label{eqh}
  \hat H_{\rm bh}=-\Omega\hat{J_x}+U\hat{J_{z}}^2,
\end{align}
where the collective spin operators are
\begin{align}\label{eq.replace}
  \hat J_\xi=\frac12\sum_{k=1}^N\hat{\sigma}_{\xi}^{(k)}.
\end{align}
The spectrum of this operator is spanned by the symmetrized states $\ket{N-n,n}$ of $n$ spins in the $-1$ eigen-state of (for instance) the $z$-component Pauli operators, and the remaining $N-n$, 
in the $+1$ eigen-state. 
Hence the dimensionality of the Hilbert space is: $\dim\mathcal H^{({\rm sym})}_N=N+1$ rather than: $\dim\mathcal H_N=2^N$ for a generic collection of $N$ spins. 
Therefore any state vector $\ket\psi$ in this symmetric subspace can be expressed as
\begin{align}\label{eq.psi.decomp}
  \ket\psi=\sum_{n=0}^N\psi_n\ket {N-n,n},\ \ \ \mathrm{with}\ \ \ \sum_{n=0}^N\modsq{\psi_n}=1.
\end{align}
To extract the crucial properties of the eigen-states of the Hamiltonian from Eq.~\eqref{eqh} we follow the steps of Ref.~\cite{ZP2008}. First, we  project the stationary Schr\"odinger equaiton
onto the $n$'th element of the basis, i.e, $\bra {N-n,n}\hat H\ket\psi=E\psi_n$, giving
\begin{align}\label{s1}
  -\frac{\Omega}{2}\left[ \psi_{n+1}\sqrt{(N-n)(n+1)}+\psi_{n-1}\sqrt{(N-n+1)n}\right]+\dfrac{U}{4}\left(N- 2n\right)^2\psi_n=E\psi_n
\end{align}
Next, we introduce the normalized population imbalance,
\begin{align}\label{eq.def.z}
  z_n=\frac{(N-n)-n}{N}=1-\frac{2n}{N}
\end{align}
which varies from $-1$ to $1$ with the increment equal to $\Delta z=2/N$. 
\begin{figure}[t!]
  \centering
  \includegraphics[width=0.9\linewidth]{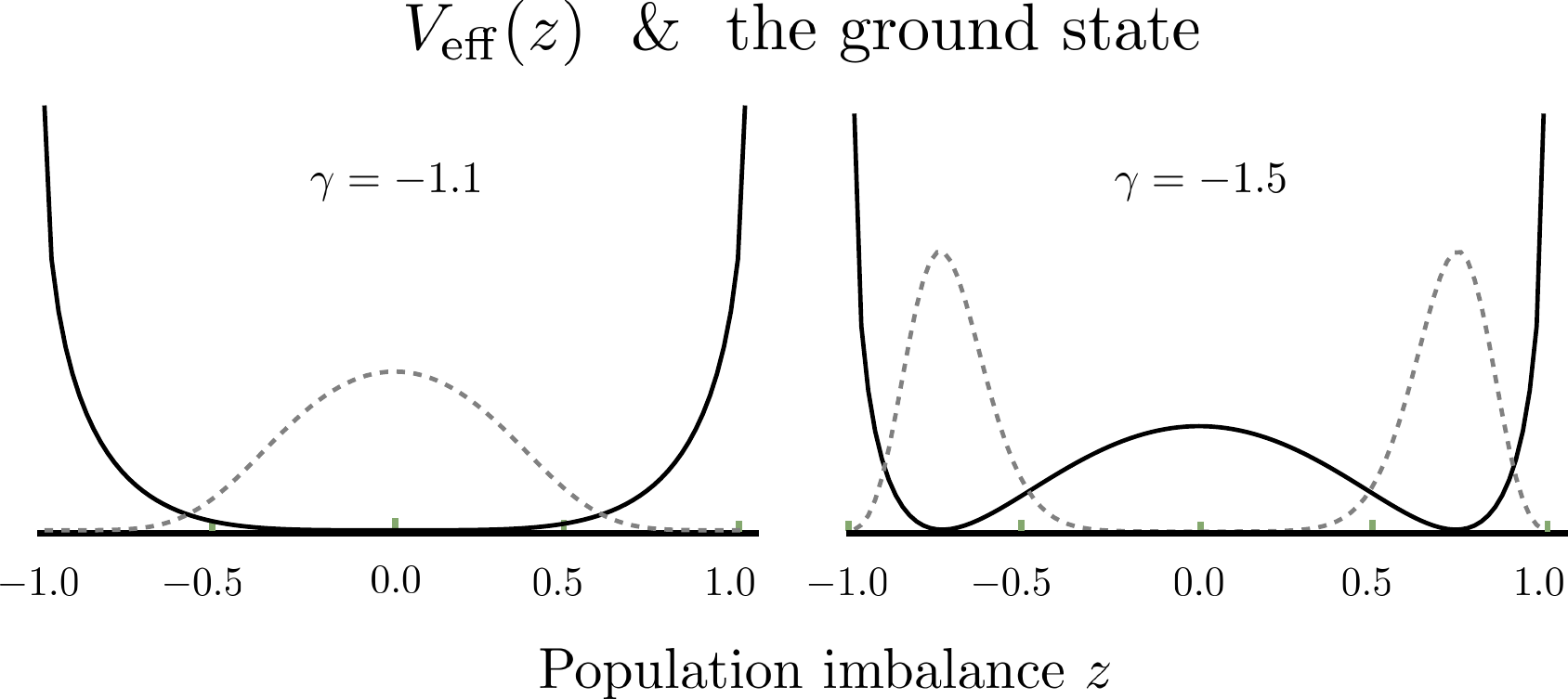}
  \caption{The effective potential $V_{\rm eff}(z)$ (solid black lines) for $N=100$ and $\gamma=-1.1$ (left) and $\gamma=-1.5$ (right). The dashed gray lines show the ground state for each case.}
  \label{fig.pot}
\end{figure}
The Eq.~\eqref{s1}, expressed in terms of this new variable becomes
\begin{align}
  &-\frac{\Omega N}{2}\left[\psi_{n+1}f_+(z_n)+\psi_{n-1}f_-(z_n)\right]+\frac{U N^2}{4}\psi_n z^{2}_{n}=E\psi_n,\\
  &\mathrm{where}\ \ \ f_\pm(z_n)=\sqrt{\frac{1\pm z_n}{2}\left( \frac{1\mp z_n}{2}+\frac{1}{N}\right)}\nonumber.
\end{align}
When $N$ is large, the $1/N$ term in $f_{\pm}$ can be neglected. Furthermore, as the increment of the $z_n$ diminishes, the discrete variable can be approximated with a continuous
$z$. In particular, this means that the finite difference tends to
\begin{align}
  \frac{\psi_{n+1}+\psi_{n-1}-2\psi_n}{(\Delta z)^2}\approx \frac{d^2}{dz^2}\psi(z)
\end{align}
In such large-$N$ regime we obtain a stationary 1D Schr\"odinger-like 
equation for a fictious unit-mass particle subject to an external potential 
\begin{align}\label{eq.pot}
  V_{\rm eff}(z)=-\sqrt{1-z^2}+z^2\gamma/2 ,
\end{align} 
with $\gamma=U N/\Omega$, i.e., 
\begin{align}\label{eql}
  \left(-\frac{2}{N^2}\sqrt{1-z^2}\frac{d^2}{dz^2} +V_{\rm eff}(z)\right)\psi(z)=\tilde E\psi(z).
\end{align}
The normalized energy is $\tilde E=\frac{2E}{\Omega N}$.
Note also that here $2/N$ is the dimensionless
equivalent of the reduced Planck constant $\hbar$. 
For the mathematical details of this derivation with the discussion of all its limitations, we refer the readers to Ref.~\cite{ZP2008}. 

The ground state of Eq.~\eqref{eql} depends crucially on the value of $\gamma$, hence on the ratio (and the sign) of the interaction-to-tunneling energies. For large and positive $\gamma$'s the 
$V_{\rm eff}(z)\eqsim\frac12\gamma z^2$, hence the problem simplifies to that of the 1D harmonic potential. As $\gamma$ grows, the width of the ground state Gaussian  shrinks. This
implies the diminishing fluctuations of the population imbalance, that, since the intra-mode coherence is mostly maintained, signals the spin-squeezing of the state~\cite{kitagawa1993squeezed,wineland1994squeezed}.
\begin{figure}[t!]
  \centering
  \includegraphics[width=0.6\linewidth]{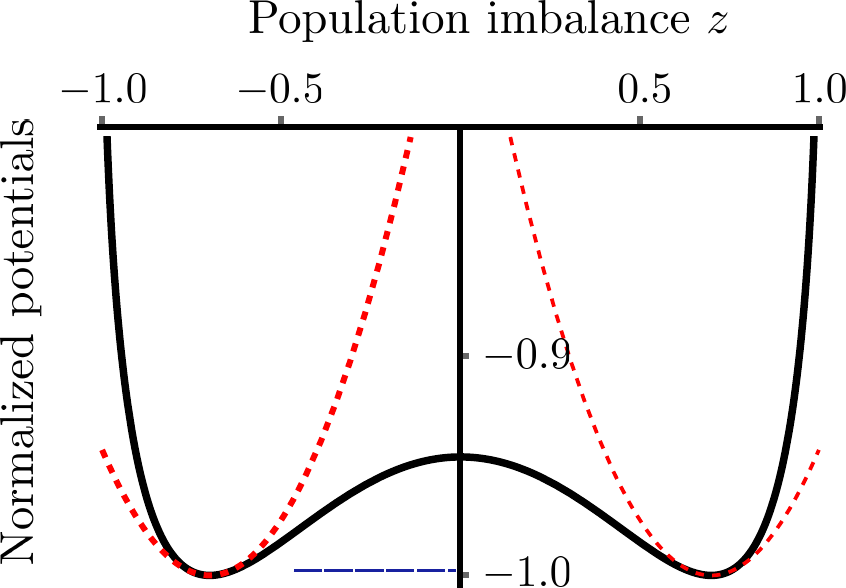}
  \caption{The comparison of the potentials $V_{\rm eff}(z)$ (normalized to $V_{\rm min}=-1$) from Eq.~\eqref{eq.pot} and from Eq.~\eqref{eq.pot.harm} for $N=500$ and $\gamma=-1.4$. The horizontal dashed blue line shows
    the energy scales of the few lowest-lying states of either potential.}
  \label{fig.pots}
\end{figure} 

On the negative-$\gamma$ side, the shape of the ground state is more diversified, depending on the value of the parameter. 
For $\gamma\gtrsim\gamma_0\equiv-1$ the wave-function is still a Gaussian, but now its width is growing as $\gamma$ approaches $\gamma_0$---this in turn signals the phase-squeezing of the sample~\cite{chwed_njp}. 
However, when the $\gamma_0$ is crossed, $V_{\rm eff}(z)$ changes abruptly indicating the passage through the quantum phase transition (QPT).
This is visualized in Fig.~\ref{fig.pot} where we show how the
potential and the ground state change across this highlighted point. Upon crossing $\gamma_0$, the potential breaks and develops two-minima at positions $\pm z_0$ with
$z_0=(1-1/\gamma^2)^{1/2}$. 
A quantum state that respects the left/right
symmetry of the problem is a macroscopic superposition of two separated wave-packets. The fluctuations of the population imbalance rapidly grow and 
as $\gamma\rightarrow-\infty$ the maxima of $\psi(z)$ separate tending towards the NOON state, which in the ket notation used in Eq.~\eqref{eq.psi.decomp} is $\ket{\psi} = 1/ \sqrt{2}\big(\ket {N,0}+\ket {0,N}\big)$.

The shape of the potential after the breaking suggests that some of the properties of the state vector can be extracted by locally approximating the $V_{\rm eff}(z)$ 
with two harmonic oscillators located at $\pm z_0$, namely
(recall that the fictitious particle has a mass set to unity)
\begin{align}\label{eq.pot.harm}
  V_{\rm eff}(z)\simeq\frac12\omega^2(z\pm z_0)^2+V_0,
\end{align}
where $\omega=\sqrt{\gamma(1-\gamma^2)}$ and $V_0=\frac{\gamma^2+1}{\gamma}$. In Fig.~\ref{fig.pots} we show how this approximation works for $N=500$ qubits and $\gamma=-1.4$. Roughly speaking, the 
position of the lowest-lying levels, denoted by the horizontal blue dashed line seems to be deep in the regime where the approximation works. Nevertheless, this must be
analyzed in more detail, as discussed below.

\begin{figure}[t!]
  \centering
  \includegraphics[width=0.9\linewidth]{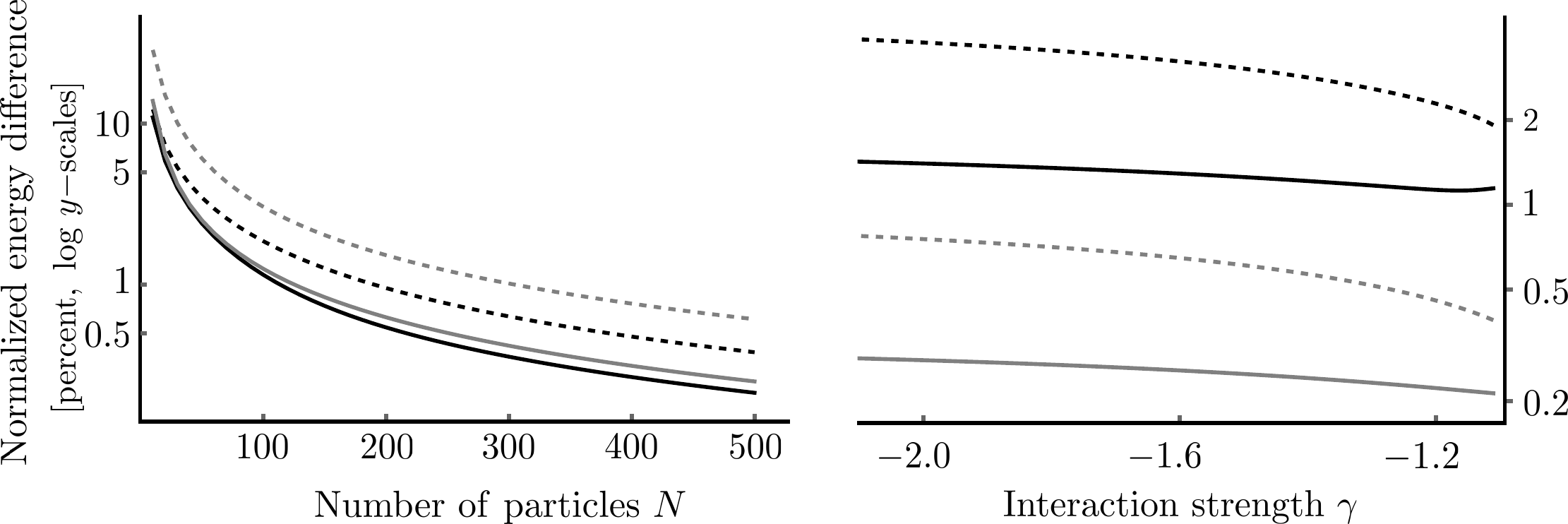}
  \caption{Left: The normalized energy difference, $\Delta E_i$ see Eq.\eqref{eq.ratio}, for $i=0$ (the ground state, solid lines) and $i=1$ (the first excited state, dashed lines) 
    of the potentials~\eqref{eq.pot} and~\eqref{eq.pot.harm} for $\gamma=-1.1$ (black) and $\gamma=-1.5$ (gray) as a function of $N$.
    Right: Here, $N$ is fixed to $N=100$ (black) and $N=500$ (gray), while $\gamma$ is varied.}
  \label{fig.ener}
\end{figure}

Note that the harmonic potential~\eqref{eq.pot.harm} is an approximation of an already approximate model of the Schr\"odinger-like equation from Eq.~\eqref{eql}. In order to make sure that in this process
the errors do not accumulate to some unacceptable level, we perform the exact diagonalization of the BH Hamiltonian from Eq.~\eqref{eqh} and compare its two lowest-lying energy levels
with those coming from the harmonic approximation (HA), 
see Eq.~\eqref{eq.pot.harm}. In Fig.~\ref{fig.ener} we show the normalized energy difference between these two outcomes, expressed in percent units, namely 
\begin{align}\label{eq.ratio}
  \Delta E_i=\left|\frac{E_i^{\rm(BH)}-E_i^{\rm(HA)}}{E_i^{\rm(BH)}}\right|\times100\%
\end{align}
for $i=0$ (ground state) and $i=1$ (first excited state). 
The left panel shows the difference for two $\gamma$'s equal to -1.1 and -1.5 as a function of $N$. As expected from the procedure described above, the discrepancy diminishes as $N$ grows and for $N=500$
can be safely kept below 1\%. Similarly, the variation of $\gamma$ for fixed $N$'s (100 and 500), see the right panel, confirms the satisfactory precision of the harmonic approximation.

\begin{figure}
  \centering
  \includegraphics[width=1.0\textwidth]{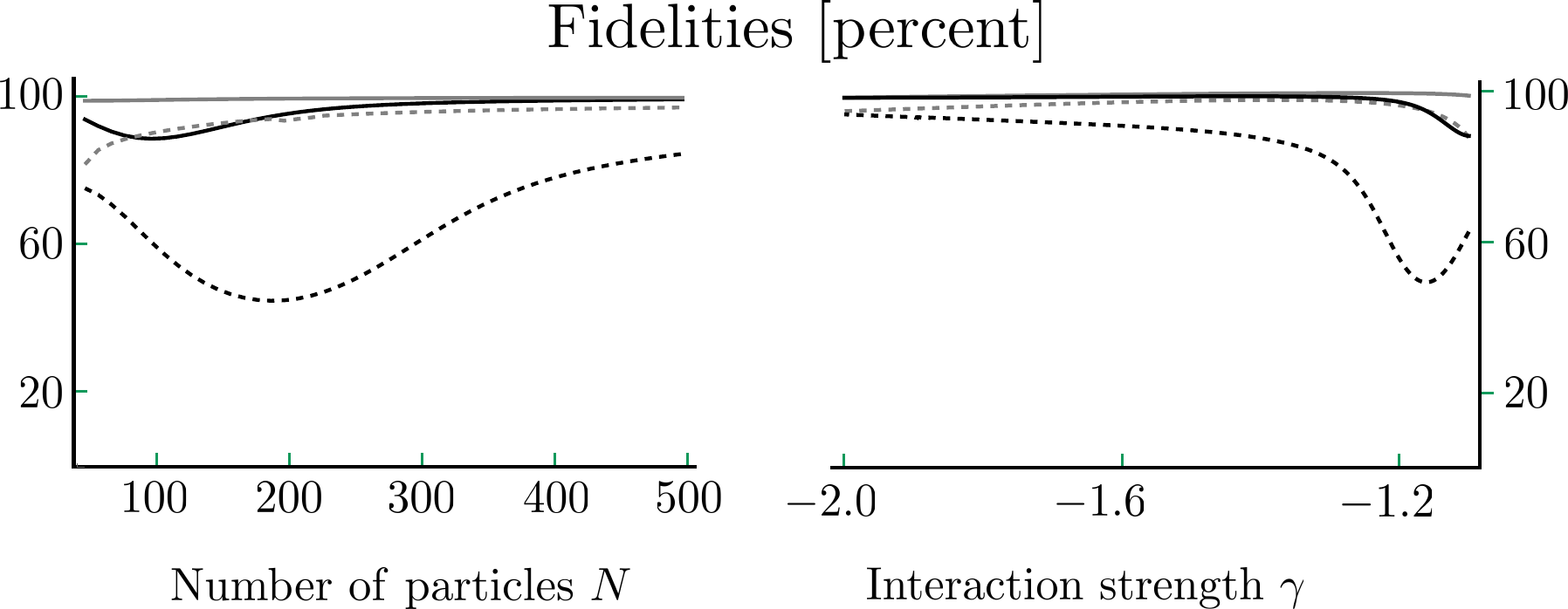}
  \caption{The quality of the harmonic approximation quantified by the fidelity $\mathcal F_i$ of the ground- and first-excited state, expressed in percent, see Eq.~\eqref{eq.fid} 
  The left panel displays $\mathcal F_i$ as a function of the number of particles $N$ for $\gamma=-1.1$ (black) and $\gamma=-1.5$ (gray). The right panel is the reverse: here $N$'s are
  fixed to $100$ (black) and 500 (gray) while $\gamma$ changes.}
  \label{fig.fid}
\end{figure}

Last but not least, we scrutinize the quality of the harmonic approximation by calculating the overlap between the eigen-states corresponding to these two energy levels with the eigen-states of
the exact BH Hamiltonian
\begin{align}\label{eq.fid}
  \mathcal F_i=\modsqs{\braket{\psi_i^{({\rm BH})}}{\psi_i^{({\rm HA})}}}\times100\%\ ,
\end{align}
again with $i=0,1$. In Fig.~\ref{fig.fid} we show that for the ground state, the fidelity $\mathcal F_0$ is above 90\% for a wide range of $\gamma$'s and $N$'s. While the fidelity $\mathcal F_1$
can be unsatisfactory for smaller $N$'s and in the direct vicinity of the QPT, we will argue that even in $T\neq0$ this is not a concern---the Bell correlator we discuss below
is mostly characterized by the properties of the ground-state.

In this section we have shown how to replace to exact quantum BH model with an approximate twin-harmonic approximation. It is valid at the negative side of the quantum critical point and will be
our working horse for the derivation of simple yet powerful analytical formul\ae\ of the many-body Bell correlations. 

\section{Many-body Bell correlator}

We now briefly review the theory behind the method of detecting many-body Bell correlators that will be used in this work. For an extensive discussion of its properties, 
see References~\cite{cavalcanti2007bell,he2011entanglement,cavalcanti2011unified,spiny.milosz,JC2022}. 

Consider $m$ objects, each being a subject of local measurements of two binary quantities 
$\sigma_x^{(k)}=\pm1$ and $\sigma_y^{(k)}=\pm1$ with $k=1\ldots m$. These outcomes are combined locally (for each object) 
into $\sigma_+^{(k)}=\frac12(\sigma_x^{(k)}+i\sigma_y^{(k)})$ and the following correlator is constructed
\begin{align}\label{eq.corr}
  \mathcal E_m=\modsq{\av{\sigma_+^{(1)}\ldots\sigma_+^{(m)}}}.
\end{align}
Here, the $\av{\cdot}$ denotes averaging over experimental repetitions. If the $\mathcal E_m$ can be reproduced by system that is consistent with the 
local hidden variable theory~\cite{epr,Bell,bell_rmp,bell_local}, then
this average can be expressed in terms of an integral over this (possibly multi-variate) variable $\lambda$ distributed with a probability density $p(\lambda)$, namely
\begin{align}
  \mathcal E_m=\modsq{\int\!\!d\lambda\,\,p(\lambda)\,\sigma_+^{(1)}(\lambda)\ldots\sigma_+^{(m)}(\lambda)}.
\end{align}
Using the Cauchy-Schwarz inequality for complex integrals we obtain
\begin{align}\label{eq.bell}
  \mathcal E_m\leqslant\int\!\!d\lambda\,\,p(\lambda)\,\modsq{\sigma_+^{(1)}(\lambda)}\ldots\modsq{\sigma_+^{(m)}(\lambda)}=2^{-m}.
\end{align}
Thus $\mathcal E_m\leqslant 2^{-m}$ is the $m$-body Bell inequality.

This inequality can be tested with quantum systems of $m$ qubits. In such case, the correlator $\mathcal E_m$ from Eq.~\eqref{eq.corr} is replaced by its quantum-mechanical ($q$) equivalent
\begin{align}\label{eq.corr.qm}
  \mathcal E^{(q)}_m=\modsq{\tr{\hat\varrho\,\hat\sigma_+^{(1)}\otimes\ldots\otimes\sigma_+^{(m)}}}.
\end{align}
For the case of $m$-qubit Greenberg-Horne-Zeilinger (GHZ) state~\cite{greenberger1989going}, i.e., 
\begin{align}
  \ket\psi=\frac1{\sqrt2}(\ket\uparrow^{\otimes m}+\ket\downarrow^{\otimes m}),\ \ \ \mathrm{where}\  \hat\sigma_+\ket\downarrow=\ket\uparrow,
\end{align}
we obtain $\mathcal E^{(q)}_m=1/4$, hence the exponential (as a function of $m$) breaking of the bound~\eqref{eq.bell}.

This quantum correlator~\eqref{eq.corr.qm} can be adapted to symmetric systems, where only collective operations are allowed. In this case, the single-body particle-resolving operators
$\hat\sigma_+^{(k)}$ are replaced with its (still one-body) collective equivalent
\begin{align}
  \hat\sigma_+^{(k)}\longrightarrow\hat J_+=\sum_{k=1}^N\hat\sigma_+^{(k)},
\end{align}
in analogy to Eq.~\eqref{eq.replace}. Upon replacing the $\sigma_+^{(k)}$'s in Eq.~\eqref{eq.corr.qm} by $\hat J_+^m$ one notices that there are $N!/(N-m)!$ more terms in the latter case,
the consequence of the permutational invariance of the problem. Hence the proper symmetric (the symmetrization is here denoted by the tilde symbol) $m$-body Bell inequality is 
\begin{align}\label{lcor}
  \tilde{\mathcal E}^{(q)}_m\equiv\modsq{\av{\hat J ^m_{+}}}\le \left(\frac{N!}{(N-m)!} \right)^2 2^{-m}
\end{align}
and more details on its derivation can be found in Ref.~\cite{JC2022}. It is convenient to normalize the l.h.s. of this inequality by its r.h.s. and introduce~\cite{plodzien2023generation}
\begin{align}\label{eq.def.qm}
  Q_m=\log_2\left[2^m\left(\frac{N!}{(N-m)!} \right)^{-2}\tilde{\mathcal E}^{(q)}_m\right].
\end{align}
The Bell inequality then becomes
\begin{align}
  Q_m\leqslant0.
\end{align}

It is the purpose of the remainder of this work to demonstrate that both at zero temperature ($T=0$) as well as when $T>0$, there exists an  order $m=\mu$ of the Bell correlator which is significantly above
the nonlocality bound $Q_\mu>0$ in the vicinity of the QPT. This particular $Q_\mu$, as we argue below, informs that an extensive (i.e., growing linearly with  $N$) number of
qubits is Bell-correlated. Moreover, $Q_\mu$ turns out to be directly linked to how far $\gamma$ is from the quantum critical point. Benefiting from the above-verified harmonic approximation we will 
connect the correlation order $\mu$ and the value of the corresponding correlator, $Q_\mu$, with $\gamma$ using simple but versatile analytical formul\ae, and extend the analysis to non-zero
temperatures.

\begin{figure}[t!]
  \centering
  \includegraphics[width=0.6\linewidth]{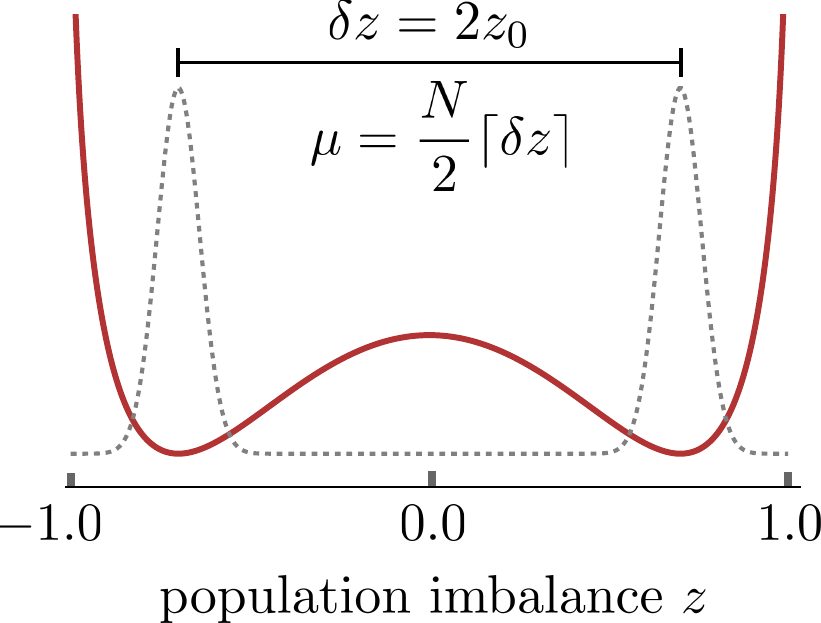}
  \caption{The ground state of the Bose-Hubbard Hamiltonian and the effective potential $V_{\rm eff}(z)$ (both in arbitrary units picked only for illustration) for $N=500$ qubits and $\gamma=-1.4$. 
    As argued in the text, the distance between the peaks determines the relevant order of the Bell correlator.}
  \label{fig.deltaz}
\end{figure}

The concept relies on the following observation. If the operator $\hat J ^m_{+}$ that determines the correlator $Q_m$ acts on a ket $\ket{N-n,n}$ it gives
\begin{align}\label{eq.corr.dens}
  \hat J ^m_{+}\ket {N-n,n}=j_{nm}\ket {N-n+m,n-m}\ \ \ \mathrm{with}\ \ \ j_{nm}=m!\sqrt{\binom{n}m\binom{N-n+m}{m}}.
\end{align}
Hence the average of this operator is equal to
\begin{align}\label{eq.sum.jm}
  \av{\hat J ^m_{+}}=\sum_{n=0}^{N-m}\varrho_{n,n+m}j_{nm},
\end{align}
i.e., a coherent sum of all the elements of the density matrix distanced by $m$ from the diagonal and weighted with $j_{nm}$'s. 
Here we used the general expression for the density matrix in the $N$-qubit symmetric subspace, namely
\begin{align}
  \hat\varrho=\sum_{nn'=0}^N\varrho_{nn'}\ketbra {N-n,n}{N-n',n'}.
\end{align}
Due to the characteristic twin-peak structure of the ground state, one $m$ stands out, namely such that is directly linked to the separation of the two maxima $\delta z=2z_0$ by means of
Eq.~\eqref{eq.def.z}, i.e., 
\begin{align}\label{eq.dm}
  \mu=\frac N2\lceil\delta z\rceil=N\lceil z_0\rceil,
\end{align}
see Fig.~\ref{fig.deltaz}. Here, $\lceil x\rceil$ denotes the integer, which is closest and bigger equal to $x$ (the ``ceil'' function). Hence the conjecture: one particular 
order of Bell correlator, at a given $\gamma$, stands out, i.e., such for which $m=\mu$ holds. 
Moreover, its
value can be reproduced with excellent precision by reducing the sum in Eq.~\eqref{eq.sum.jm} merely to the contribution coming from the peak values at these two maxim\ae\  positioned 
at $n_\pm=(N\pm\mu)/2$. 
In other words, using Eqs~\eqref{lcor} and~\eqref{eq.corr.dens}, the correlator is approximated by
\begin{align}\label{eq.qm.app}
  Q_{\mu}\simeq\log_2\left[2^{\mu}\left(\frac{N!}{(N-\mu)!} \right)^{-2}\modsq{\varrho_{n_+,n_-}j_{n_+,n_-}}\right]
\end{align}
We now show, that this very simple formula works exceptionally well for a pure ground state and allows to determine the critical temperature, above which the Bell correlator $Q_\mu$ drops below the Bell
limit.

It should be underlined that $Q_\mu$ does not provide a complete information about Bell nonlocality in the system. At fixed $\gamma$, there are various correlation orders that give $Q_m>0$. Also, there
are other ways to construct the many-body Bell correlator~\cite{zukowski2002bell}, that in principle could yield more information about the nonlocal correlations in this case. 
Nevertheless the analysis presented here has two strong points---it is simple, leading to analytical predictions, and it allows to lower-bound the extend of Bell correlations over the system, 
an important issue for many applicational aspects, such as quantum-enhanced metrology~\cite{PhysRevLett.126.210506}.  

\section{Bell correlations at $T=0$}
The Harmonic approximation yields the ground state of the system to be
\begin{align}
  \ket\psi\simeq\left(\frac\omega{2\pi N}\right)^{\frac14}\sum_{n=0}^N\left(e^{-(\frac{N-\mu}2-n)^2\frac\omega N}+e^{-(\frac{N+\mu}2-n)^2\frac\omega N}\right)\ket {N-n,n}.
\end{align}
Hence by taking the peak-values of the state coefficients [i.e., located at $n_\pm=(N\pm\mu)/2$], and substituting this result into Eq.~\eqref{eq.qm.app}, we obtain
\begin{align}
  Q_{\mu}=\log_2\left[2^{\mu}\binom{\frac{N+\mu}2}{\mu}^2\binom{N}{\mu}^{-2}\frac\omega{2\pi N}\right].
\end{align}
Keeping the dominant terms that scale with $N$, we obtain a simple but powerful expression that allows to lower-bound the strength of $\mu$-body Bell correlations in the vicinity of the
quantum critical point, namely
\begin{align}
  Q_{\mu}\simeq \mu\log_2\left[\frac{N+\mu}{N-\mu}\right]-\mu-N\log_2(\gamma^2).
\end{align}
Finally, by using Eq.~\eqref{eq.dm}, we obtain a compact formula
\begin{align}\label{eq.qm.fin}
  Q_{\mu}=Nf(\gamma),\ \ \ \mathrm{where}\ \ \ f(\gamma)=\frac{\sqrt{\gamma^2-1}}{|\gamma|}\left[2\log_2\left(\sqrt{\gamma^2-1}+|\gamma|\right)-1\right]-2\log_2|\gamma|.
\end{align}
Most importantly, the correlator is extensive in $N$, which has profound consequences for the quantitative characterization of many-body Bell nonlocality. 
Moreover, this form of $Q_\mu$ implies that the value of $\gamma$ at which the many-body Bell correlations emerge is universal (i.e., independent of $N$) and equal to $\gamma_0\simeq-1.3$. 

Note that the value of $Q_\mu$ carries information about the nonlocality depth, i.e., about how many qubits are Bell-correlated. And so, if
\begin{align}
  \mu-2-(k+1)<Q_\mu\leqslant \mu-2-k,
\end{align}
where $k\in\mathbb N$, then this correlator can be reproduced with a state of $m$ qubits, where not more that $k-2$ of them are {\it not} Bell-correlated with the 
rest~\cite{spiny.milosz,PhysRevLett.129.250402,Plodzie2023generation}. 
By using Eq.~\eqref{eq.qm.fin} we notice that the depth of Bell correlations in the ground-state can be lower-bounded as
\begin{align}
  k\leqslant N(\lceil z_0\rceil-f(\gamma))-2.
\end{align}
Hence we conclude that the number of Bell correlated qubits is an extensive function of $N$, starting from the critical point $\gamma=\gamma_0$. This confirms that $Q_\mu$ is
a useful tool to lower-bound the nonlocality depth in this system.

\section{Quantum Correlations at T>0}
Next, we calculate the Bell correlator and the nonlocality depth in non-zero temperatures, considering the thermal density matrix
\begin{align}\label{eq.st.th}
  \hat\varrho=\frac1{\mathcal Z}\sum_n\ketbra{\psi_n}{\psi_n}e^{-E_n\beta},
\end{align}
where, $\beta^{-1}=k_bT$, $k_b$ is the Boltzmann constant, $\mathcal Z$ is the statistical sum and $\hat H_{\rm bh}\ket{\psi_n}=E_n\ket{\psi_n}$ with the Hamiltonian from Eq.~\eqref{eqh}.

\begin{figure}
  \centering
  \includegraphics[width=1.0\textwidth]{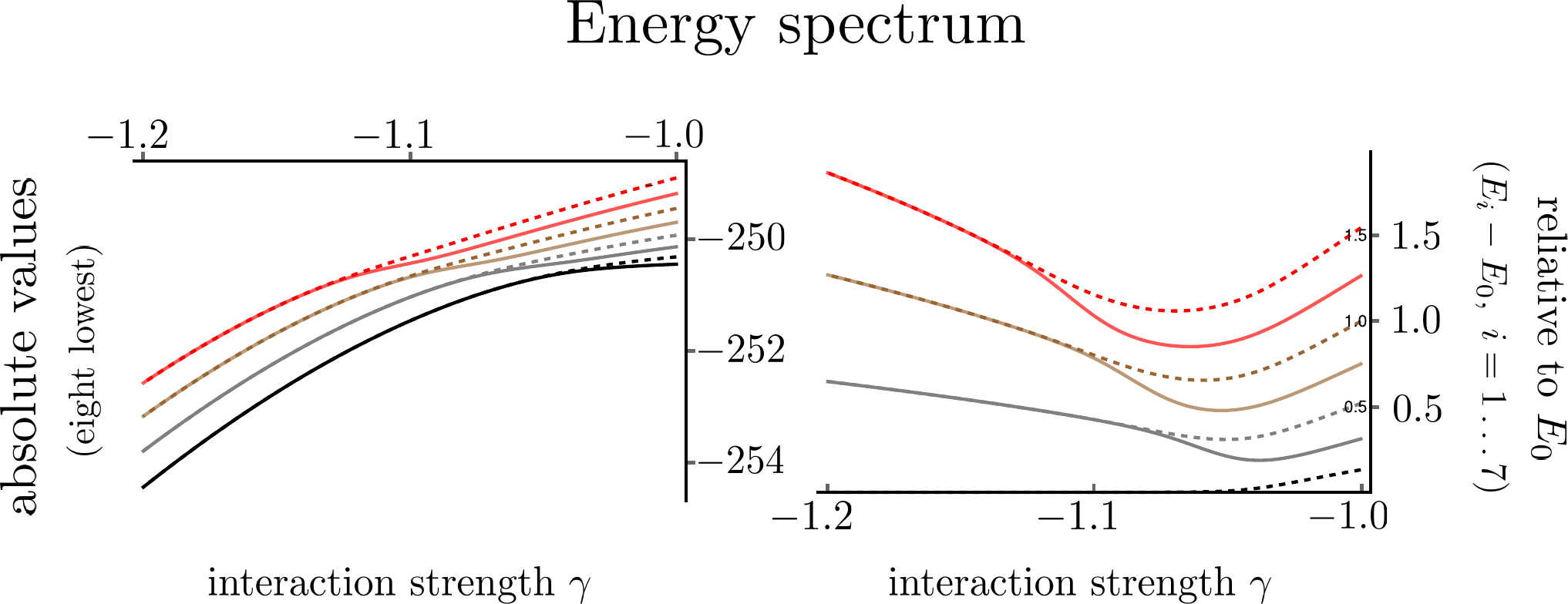}
  \caption{The energy spectrum (8 lowest levels) of the Hamiltonian $H_{\rm bh}$ for $N=500$ atoms in the vicinity of the critical point. The left panel shows the energies $E_i$ ($i=0\ldots7$) while
  the right panel displays the seven energy gaps with respect to the ground state $E_0$.}
  \label{fig.spectr}
\end{figure}

The characteristic temperature scales are set by the eigen-energies (with respect to the ground state) of the Hamiltonian, hence first we plot a part of the spectrum for $\gamma$'s of interest and $N=500$ atoms. see Fig.~\ref{fig.spectr}. The energy gaps $\delta_i=E_i-E_0$ of the seven lowest-lying excited states are all of the order of unity, apart from the first one, i.e., $\delta_1$, which
quickly converges to 0 upon passing the QPT point. This observation allows us to identify the relevant scale of $\beta$'s.

\begin{figure}
  \centering
  \includegraphics[width=1.0\textwidth]{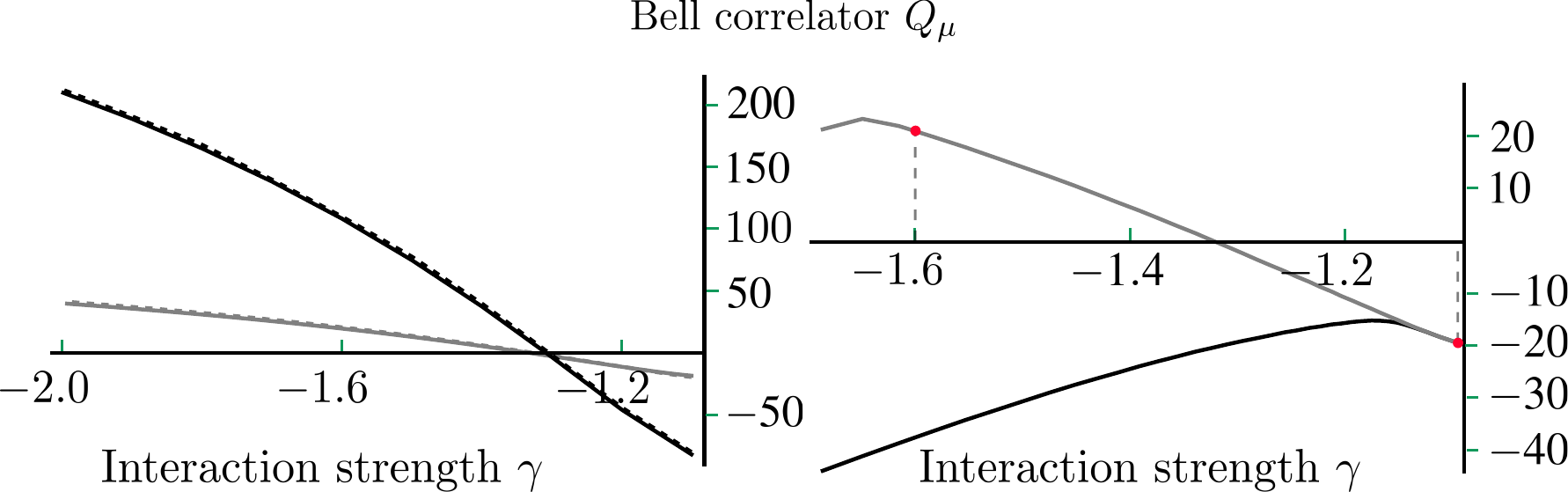}
  \caption{The Bell correlator calculated with the full Bose-Hubbard Hamiltonian (solid lines) and with the approximate formul\ae\ (dashed lines). Left: $T=0$ for $N=100$ (gray) and $N=500$ (black) qubits.
    The approximate formula is in Eq.~\eqref{eq.qm.fin}. Right: the correlator $Q_\mu$ calculated with the thermal state, see Eq.~\eqref{eq.st.th} with $N=100$ qubits. The 
    black solid line is for $k_bT_1=10\%\times\delta_1$, which is the gap taken at $\gamma=-1.1$. The gray solid line is for $k_bT_2=10\%\times\delta_1$, where the gap is for $\gamma=-1.6$. 
    Hence $T_2>T_1$. The vertical dashed lines and red dots indicate those values of $\gamma$. Clearly, soon upon passing these points, when the temperature becomes comparable to the corresponding
    gap, the correlator drastically drops.
  }
  \label{fig.temp}
\end{figure}

A following toy model indicates that the temperature should be kept well below this smallest gap, $\delta_1$, to retain high values of $Q_\mu$. Let's approximate the twin-peak ground 
state with what is sometimes called a ``Schr\"odinger kitten'' state
\begin{align}
  \ket{\psi_0}=\frac1{\sqrt2}\left(\ket{\frac{N+\mu}2,\frac{N-\mu}2}+\ket{\frac{N-\mu}2,\frac{N+\mu}2}\right).
\end{align}
The gap $\delta_1$ dropping to zero is the result of the fact that this state is almost degenerate with
\begin{align}
  \ket{\psi_1}=\frac1{\sqrt2}\left(\ket{\frac{N+\mu}2,\frac{N-\mu}2}-\ket{\frac{N-\mu}2,\frac{N+\mu}2}\right).
\end{align}
This is just an approximation of the true ground and first-excited states at finite $\gamma$. The exact degeneracy is when $\gamma\rightarrow\infty$ (i.e., when $\mu\rightarrow N$). 
Nevertheless, when the temperature is kept sufficiently low so that only $\ket{\psi_0}$ and $\ket{\psi_1}$ are populated, the thermal density matrix is approximately
\begin{align}
  \hat\varrho\simeq\frac1{1+e^{-\delta_1\beta}}\left(\ketbra{\psi_0}{\psi_0}+\ketbra{\psi_1}{\psi_1}e^{-\delta_1\beta}\right),
\end{align}
Substituting this $\hat\varrho$ into Eq.~\eqref{eq.def.qm} gives
\begin{align}
  Q_\mu=\log_2\left(2^\mu\modsq{\frac12\binom{N}{\mu}^{-1}\binom{\frac{N+\mu}2}{\mu}\frac{1-e^{-\delta_1\beta}}{1+e^{-\delta_1\beta}}}\right).
\end{align}
Clearly, when $\beta\ll\delta_1$, i.e., when $k_bT$ is much larger than the gap, the correlator vanishes. This conjecture is fully confirmed by
the exact diagonalization, that gives the thermal state and the correlator $Q_\mu$, see the right panel of Fig~\ref{fig.temp}. Here. the temperature is picked to be 
equal to 10\% of the energy gap $\delta_1$, value of which is calculated either using $\gamma=-1.1$ (solid black line) or $\gamma=-1.6$ (solid gray line). The correlator $Q_\mu$ drops drastically when the
interaction strength $\gamma$ is far away from the QPT so that the population of the excited state becomes significant.  

Recall that the many-body Bell correlations detected by $Q_\mu$ require $\gamma<-1.3$, see the discussion below Eq.~\eqref{eq.qm.fin}. This observation, together with the results from the
above paragraph, yield the critical temperature, above which the Bell correlations will not be detected by $Q_\mu$. Namely, if on one hand, $\gamma$ must be below $\gamma_0$ and the temperature must be
at least one order of magnitude smaller than the energy gap, the maximal temperature  at which $Q_\mu>0$ is set by a fraction of the energy gap $\delta_1$ at $\gamma_0$. Since this gap rapidly shrinks
with growing $N$, we conclude that the many-body Bell correlator $Q_\mu$ is the more sensitive to any thermal excitations, the higher $N$ is.

\section{Conclusion}

In this work  we have analyzed the character of the many-body Bell correlations in an interacting multi-qubit systems with particle-exchange symmetry, such as the fully connected
Ising model in a perpendicular magnetic field or the tight-biding Bose-Hubbard Hamiltonian of a Bose-Einstein condensate in a double-well potential.
Our analysis resides on the observation that
such systems can be mapped onto an effective Schr\"odinger-like equation. This allows for detailed analytical calculations, using a harmonic approximation of the local minima of the
effective potential. As our main result, we show that upon passing a quantum phase transition, the ground state of the system is characterized by strong Bell correlations that are extensive in $N$.
We also show the existence of the threshold temperature, above which, once the thermal noise dampens the Bell correlations. 

In the expected advent to NISQ devises, such precisely tailored analysis might help in solving  various theoretical and experimental problems.

\section*{Acknowledgements}

This project was funded by the National Science Centre, Poland, within the QuantERA II Programme that has received funding from the European Union’s Horizon 2020 research and innovation
programme under Grant Agreement No 101017733, Project No. 2021/03/Y/ST2/00195.

\section*{ORCID IDs}

Danish Ali Hamza: \url{https://orcid.org/0000-0001-9829-8260}\\
Jan Chwede\'nczuk: \url{https://orcid.org/0000-0002-9250-4227}

\section*{References}

\providecommand{\newblock}{}

\end{document}